\documentclass[prl,twocolumn,superscriptaddress,showpacs,nobalancelastpage]{revtex4}
\usepackage{graphicx}

\begin{document}

\title{Induced superconductivity in the three-dimensional topological insulator HgTe}

\author{Luis Maier}
\affiliation{Physikalisches Institut (EP3), University of
W\"{u}rzburg, Am Hubland, D-97074, W\"{u}rzburg, Germany}

\author{Jeroen B. Oostinga}
\email{jeroen.oostinga@physik.uni-wuerzburg.de}
\affiliation{Physikalisches Institut (EP3), University of
W\"{u}rzburg, Am Hubland, D-97074, W\"{u}rzburg, Germany}

\author{Daniel Knott}
\affiliation{Physikalisches Institut (EP3), University of
W\"{u}rzburg, Am Hubland, D-97074, W\"{u}rzburg, Germany}

\author{Christoph Br\"{u}ne}
\affiliation{Physikalisches Institut (EP3), University of
W\"{u}rzburg, Am Hubland, D-97074, W\"{u}rzburg, Germany}

\author{Pauli Virtanen}
\affiliation{Institut f\"{u}r Theoretische Physik und Astrophysik
(TP4), University of W\"{u}rzburg, Am Hubland, D-97074,
W\"{u}rzburg, Germany}

\author{Grigory Tkachov}
\affiliation{Institut f\"{u}r Theoretische Physik und Astrophysik
(TP4), University of W\"{u}rzburg, Am Hubland, D-97074,
W\"{u}rzburg, Germany}

\author{Ewelina M. Hankiewicz}
\affiliation{Institut f\"{u}r Theoretische Physik und Astrophysik
(TP4), University of W\"{u}rzburg, Am Hubland, D-97074,
W\"{u}rzburg, Germany}

\author{Charles Gould}
\affiliation{Physikalisches Institut (EP3), University of
W\"{u}rzburg, Am Hubland, D-97074, W\"{u}rzburg, Germany}

\author{Hartmut Buhmann}
\affiliation{Physikalisches Institut (EP3), University of
W\"{u}rzburg, Am Hubland, D-97074, W\"{u}rzburg, Germany}

\author{Laurens W. Molenkamp}
\affiliation{Physikalisches Institut (EP3), University of
W\"{u}rzburg, Am Hubland, D-97074, W\"{u}rzburg, Germany}

\date{\today}

\begin{abstract}

A strained and undoped HgTe layer is a three-dimensional topological
insulator, in which electronic transport occurs dominantly through
its surface states. In this Letter, we present transport
measurements on HgTe-based Josephson junctions with Nb as
superconductor. Although the Nb-HgTe interfaces have a low
transparency, we observe a strong zero-bias anomaly in the
differential resistance measurements. This anomaly originates from
proximity-induced superconductivity in the HgTe surface states. In
the most transparent junction, we observe periodic oscillations of
the differential resistance as function of an applied magnetic
field, which correspond to a Fraunhofer-like pattern. This
unambiguously shows that a precursor of the Josephson effect occurs
in the topological surface states of HgTe.

\end{abstract}

\pacs{73.23.-b, 73.25.+i, 72.80.Ey, 74.45.+c}

\maketitle

Topological insulators (TIs) are a recently discovered new class of
materials \cite{TI-RMP}. In two dimensions (2-D), a TI is an
insulating material with conducting helical states at its edges. The
first experimental realization of a 2-D TI was a HgTe quantum well,
in which the topological state of matter was unambiguously
demonstrated by the observation of the quantum spin Hall effect in
electronic transport experiments \cite{07-KOE}. The first
three-dimensional (3-D) TIs, Bi$_{1-x}$Sb$_x$, Bi$_2$Se$_3$ and
Bi$_2$Te$_3$, were discovered soon after \cite{3D-TI}. 3-D TIs are
bulk insulators with conducting states at their surfaces. Since
these surface states are effectively described by Dirac
Hamiltonians, the corresponding quasiparticles behave as chiral
Dirac fermions, i.e. the spin and momentum of the electrons are
locked \cite{TI-RMP}.

Recently, it has been predicted that, due to the superconducting
proximity effect, a Dirac fermion at the surface of a 3-D TI can be
split into two so-called Majorana quasiparticles \cite{08-FU}. These
are zero-energy bound states that may be present at the surface when
a TI is contacted by a superconductor. The realization of a
superconductor-topological insulator-superconductor (S-TI-S)
junction is an important first step to study the existence of
Majorana fermions. S-TI-S junctions based on Bi$_2$Se$_3$
\cite{11-SAC} and Bi$_2$Te$_3$ \cite{11-VEL} have been recently
reported and a Josephson supercurrent has been observed in both these 3-D
TIs. However, since the Bi-based 3-D TIs have a large density of
states in their bulk, only a small part of the observed supercurrent
can be attributed to superconducting correlations in the surface
states. We recently demonstrated that strained HgTe is a 3-D TI with
negligible bulk conductance when the Fermi level is inside its
bandgap \cite{11-BRU}. This material is thus a more promising
candidate to study the Josephson effect in S-TI-S junctions without
being hindered by bulk states.

In this Letter, we study electronic transport through S-TI-S
junctions based on superconducting Nb wires contacted to a strained
HgTe layer, in which the electrical conduction is dominated by
surface state transport. Although it turns out that the Nb-HgTe
interfaces have a low transparency, we clearly observe induced
superconductivity and transport through surface states approaching
the supercurrent regime. Our results provide clear evidence of
superconducting correlations in the topological surface states of
HgTe.

Two bulk HgTe layers of 50 and 70 nm thickness, respectively, are
grown by molecular beam epitaxy on a CdTe substrate. The samples are
etched in an Ar plasma to obtain 2 $\mu$m wide HgTe stripes. On the
top surface of the HgTe stripes, 68 nm thick Nb superconducting
wires are deposited by using ultra high vacuum sputtering and lift-off techniques.
Each pair of closely-spaced neighboring wires forms an S-TI-S
junction (Figs.~1a,~b). The transport measurements on these devices
have been performed at dilution refrigerator temperatures. We have
measured the differential resistance of the junctions as function of
magnetic field and current bias by using standard lock-in detection
techniques. Here we will discuss the data of two of the closest junctions with 150~nm contact spacing; other junctions show consistent behaviour.

In a first set of measurements we investigate the superconductiong properties of the 50~nm thick HgTe layer. In order to characterize the transport properties of this bulk HgTe layer, we have fabricated from the same wafer as used for one of the S-TI-S junctions, a large six-terminal Hall bar with a channel length of 600~$\mu$m and width of 200~$\mu$m (similar to the device reported in Ref. \cite{11-BRU}). From the longitudinal and
transversal magnetoresistance measurements at $T = 4.2$ K, we extract an electron density of $n_e \approx 5 \cdot 10^{11}$~cm$^{-2}$ and a mobility of $\mu_e \approx 3 \cdot 10^4$~cm$^2$/Vs. These values are typical for strained HgTe layers in which electronic transport is dominantly through their Dirac surface states \cite{11-BRU}. Moreover, at magnetic fields higher than 2 T we observe a series of quantization steps in the Hall resistance, indicating that transport is indeed occurring through 2-D states \cite{SI, 12-SHU}.

Since the transport is dominated by surface states, it is possible to investigate the behavior of a S-TI-S junction, fabricated on this same 50 nm thick layer. In order to study transport in the superconducting regime, we have performed two-terminal measurements at a base temperature of $T \approx 25$ mK (which is far below the experimentally verified critical temperature
of Nb, $T_c \approx$ 9 K). Since each Nb terminal is split into two leads $-$ one for current injection and the other for voltage measurement $-$ we can simultaneously measure the differential resistance of the junction and the electrochemical potential difference between the Nb contacts (Fig.~1b). The differential resistance ($dV/dI$) as a function of the measured voltage across the junction ($V$) provides detailed information about the energy dependence of the transmission probability of quasiparticles through the junction \cite{Tinkham}.

\begin{figure}[!t]
\includegraphics[width=3.4in]{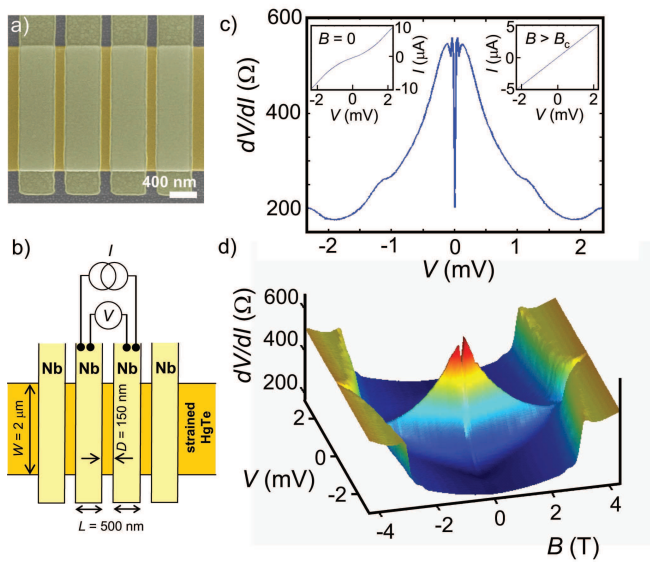}
\caption{(a) Scanning electron microscopy image of an etched HgTe
stripe contacted by four Nb wires. (b) Schematic picture of three
Nb-HgTe-Nb junctions. (c) $dV/dI$ as function of measured voltage
across one of the junctions ($B=0$). Insets show $I-V$
characteristics at $B=0$ (left) and $B>B_c$ (right). (d) $dV/dI$
versus $V$ and applied $B$.} \label{Fig1}
\end{figure}

Fig.~1c shows $dV/dI$ as function of $V$ for one of the junctions at
zero magnetic field. At $|V| \approx$ 2 mV, we observe a pronounced
dip in $dV/dI$. This is when the potential difference between both
Nb contacts equals twice the superconducting gap ($\Delta_{Nb}
\approx$ 1 meV). When $V = \pm 2\Delta_{Nb}/e$, the upper band edge
of the quasiparticle states in one Nb lead is aligned with the lower
band edge of the other (Fig.~2c). At these energies, the
quasiparticle density of states in the Nb as well as the Andreev
reflection (AR) probability is enhanced \cite{OBTK}, which gives
rise to an increased transmission probability at both interfaces,
and thus a decrease of the differential resistance of the junction
\cite{Tinkham}.

For $|V| \gtrsim 2$ mV, when the potential difference between the Nb contacts is larger than $2\Delta_{Nb}/e$, the differential resistance is $R_n \approx$ 200 $\Omega$. This is the normal conductivity transport regime, where the superconducting gaps of the Nb leads have no energy overlap and single quasiparticles from filled states below the gap in one lead are transmitted to empty states above the gap in the other lead.

If $|V| < 2\Delta_{Nb}/e$, the superconducting gaps of the Nb leads
have an energy overlap. In this sub-gap regime, quasiparticles can
only be transmitted through the junction by AR processes at the
interfaces \cite{OBTK}. Fig.~1c shows that below $|V| \approx 2$ mV
a strong increase of the differential resistance occurs on lowering
the voltage difference between the Nb leads . This indicates that
the AR probability is suppressed due to the low transparency of the
Nb-HgTe interfaces.

A further characterization of the junction is done by measuring the
magnetic field dependence of the $dV/dI$ versus $V$ characteristics
(Fig.~1d). At perpendicular magnetic fields ($B$) larger than 3 T,
we observe a steep increase in the differential resistance, which is
due to the transition of Nb from superconducting to normal state. If
the magnetic field is decreased below $B_c \approx$ 3 T, the
superconducting gap opens and the band edge can be probed by
measuring the corresponding $dV/dI$ dip at $V = \pm 2\Delta_{Nb}/e$
as function of magnetic field (Fig.~3a). This dip, where
$\frac{\partial}{\partial V}(dV/dI) = 0$, is visualized in a
colorplot by numerically differentiating $dV/dI$ versus $V$ and
plotting the obtained absolute value versus $V$ and the applied
magnetic field (Fig.~3b). The figures clearly show that the gap
increases from 0 to $\sim$1 meV when the magnetic field decreases
from $\sim$3 to 0 T.

\begin{figure}[!t]
\includegraphics[width=3.4in]{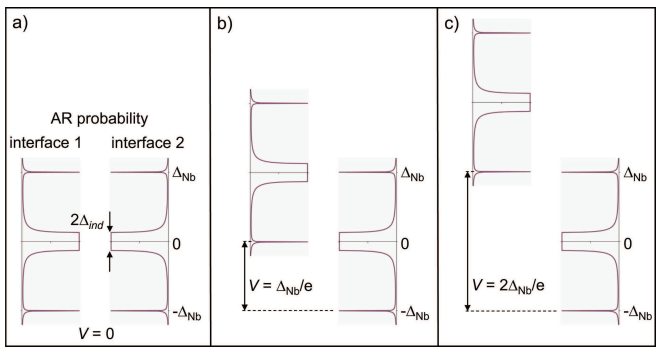}
\caption{The appearance of the main features in the characteristics
of $dV/dI$ versus $V$ can be explained by considering the AR
probabilities at both Nb-HgTe interfaces \cite{SI}. The schematic
pictures show that the transmission probability through the junction
is highest when the AR probability at both interfaces is highest:
that is for (a) $|V|=0$, (b) $|V|=\Delta_{Nb}/e$, or (c)
$|V|=2\Delta_{Nb}/e$. Note that at the band edges
($\pm\Delta_{Nb}$), the quasiparticle density of states in Nb is
enhanced, leading to an additional increase in transmission
probability at these energies.} \label{Fig2}
\end{figure}

\begin{figure}[!t]
\includegraphics[width=3.4in]{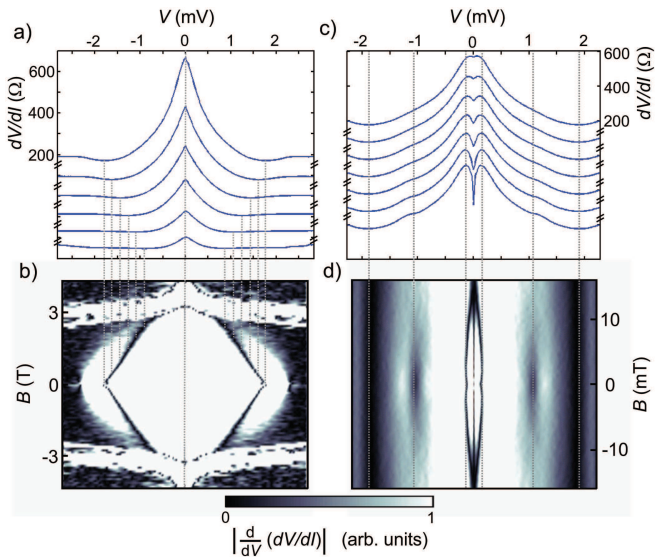}
\caption{(a) $dV/dI$ versus $V$ at high magnetic fields (starting
from top: $B=$ 0.04 T, 0.52 T, 1.00 T, 1.48 T, 2.04 T, 2.52 T; the
curves are shifted with respect to the top curve for clarity). (b)
Colorplot of the numerically derived data of $|\frac{d}{dV}(dV/dI)|$
versus $V$ and $B$. (c) $dV/dI$ versus $V$ at low magnetic fields
(starting from top: $B=$ 12.2 mT, 10.2 mT, 8.3 mT, 6.4 mT, 4.5 mT,
2.6 mT, 0.6 mT; the curves are shifted with respect to the top curve
for clarity). (d) Colorplot of the numerically derived data of
$|\frac{d}{dV}(dV/dI)|$ versus $V$ and $B$.} \label{Fig3}
\end{figure}

Interestingly, Fig.~1c shows a strong dip in $dV/dI$ at very small bias. Figs.~3c,~d show that this zero-bias anomaly (ZBA) appears at $|V| \lesssim$ 100 $\mu$V and $|B| \lesssim$ 10 mT. A ZBA can appear in Josephson junctions due to the superconducting proximity effect in the normal conductor \cite{ZBA}. Also in our S-TI-S junction, Cooper pairs "leak" from the Nb into the HgTe below the contacts, yielding electron correlations in the surface states \cite{10-STA}.These correlations give rise to induced superconductivity on an energy scale of the order of the induced superconducting gap ($\Delta_{ind}$) which depends on the transparency of the contacts. The fact that the observed ZBA is very strong indicates the surface character of the induced superconductivity in HgTe \cite{SI}. In conventional materials with bulk transport, the ZBA is usually much weaker due to the decay of the correlations with distance from the superconductor \cite{ZBA}. When the bias across the junction is smaller than twice the induced gap ($|V| < 2\Delta_{ind}/e$), the AR probability at each interface is strongly enhanced (Fig.~2a), yielding a large suppression of the differential resistance. In our junction, the ZBA appears below $|V| \approx$ 100 $\mu$V (Figs.~4a,~b), indicating an induced superconducting gap of $\Delta_{ind} \approx$ 50 $\mu$eV, which is close to a theoretically expected value for the topological surface states of HgTe \cite{SI}.

The ZBA completely disappears when a perpendicular magnetic field of
$|B| \gtrsim$ 10 mT is applied. A magnetic field gives rise to a
phase gradient in the order parameter and induces screening currents
in the Nb \cite{MovCond}, as has been previously observed in Nb/InAs
Josephson junctions \cite{09-ROH}. The moving condensate leads to
momentum and energy transfer in AR processes, resulting in a Doppler
shift of the energy of an Andreev reflected quasiparticle with
respect to the incoming quasiparticle. When this Doppler shift
becomes comparable to $\Delta_{ind}$ the ZBA will disappear
\cite{MovCond}. In our junction, this occurs at a magnetic field
value of $\sim$10 mT (Figs.~3c,~d and 4b), which is of the same order
as the estimated value obtained from calculations \cite{SI}.

When we consider the sub-gap regime of the $dV/dI$ versus $V$
characteristic in Fig.~1c more carefully, we observe a feature at
$|V| \approx$ 1 mV. Its position corresponds approximately to the
value of the superconducting gap in Nb ($\Delta_{Nb} \approx$ 1
meV). At this bias voltage, the band edge of one Nb lead is aligned
with the center of the gap of the other lead. This alignment leads
to an increase of the transmission probability through the junction
due to the enhanced AR probability at both interfaces (Fig.~2b). The
observed kink in $dV/dI$ at $V = \pm \Delta_{Nb}/e$ is therefore
closely related to the ZBA. Fig.~3c,~d show that this feature indeed
disappears at a magnetic field larger than $\sim$10 mT, similar to
the scale associated with the disappearance of the ZBA.

\begin{figure}[!t]
\includegraphics[width=2.8in]{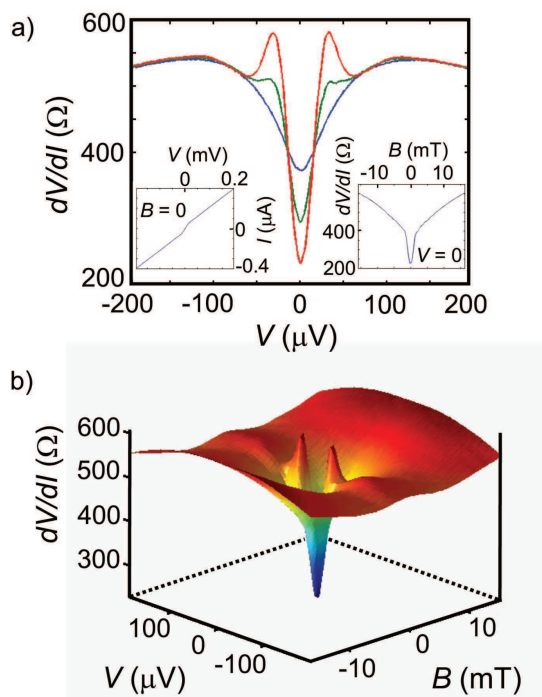}
\caption{(a) Low-bias characteristics of $dV/dI$ versus $V$ at $B=$
0 mT (red), 0.5 mT (green) and 1.0 mT (blue). Left inset shows the
$I-V$ characteristic at $B=0$. Right inset shows $dV/dI$ versus $B$
at $V=0$. (b) $dV/dI$ versus $V$ and $B$.} \label{Fig4}
\end{figure}

\begin{figure}[!t]
\includegraphics[width=2.8in]{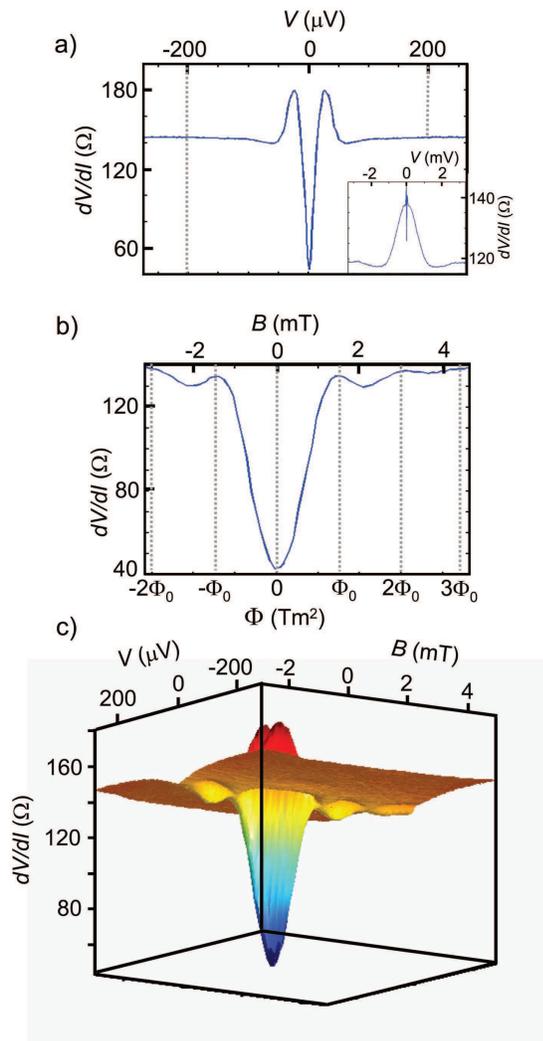}
\caption{(a) $dV/dI$ versus $V$ for a second junction with more
transparent Nb-HgTe interfaces. (b) $dV/dI$ as function of $B$ at
$V=0$. (c) $dV/dI$ as function of $V$ and $B$.} \label{Fig5}
\end{figure}

So far, we have shown that the Nb contacts induce superconductivity in this HgTe system. These electron correlations occur at least below each contact. If electron transport through the HgTe between both Nb contacts is phase coherent, the Josephson effect should give rise to bound states through which a supercurrent can flow. The critical value of such a current ($I_c$) depends on the phase difference between the condensates in both Nb layers
\cite{62-JOS}. We do not, however, observe a dissipationless current, indicating that it is destroyed by phase decoherence mechanisms. A supercurrent is absent if the thermal energy of the electromagnetic noise is larger than the Josephson energy \cite{Tinkham}, which can occur when the wires leading to the sample are not sufficiently filtered and shielded \cite{Filtering}. Given the incomplete filtering in our present set-up, this my well explain the absence of a supercurrent. However, the signature of a crossover towards a Josephson supercurrent is already visible in Figs.~4a,~b, where we observe an enhanced suppression of the differential resistance when $|V| \lesssim$ 40 $\mu$V and $|B| \lesssim$ 1 mT.

To further investigate the ZBA, we have fabricated a second array of S-TI-S junctions. The junctions have the same geometry, but the contact interfaces have been improved by a modified fabrication process. After e-beam patterning and development, we have now cleaned the HgTe surface by exposing it to a very mild, low-power, Ar plasma prior to Nb sputtering. This plasma cleaning results in more transparent Nb-HgTe interfaces without affecting the quality of the surface states. This device has been fabricated on a similar HgTe layer of 70 nm thickness \cite{SI}.  The differential resistance of this junction (see inset of Fig.~5a) has a smaller increase in the subgap region ($|V| < 2$ mV), indicating a smaller tunnel barrier at the interfaces. The ZBA appears at a higher bias value than in the previous sample (at $|V| \lesssim$ 200 $\mu V$; see dotted line in Fig.~5a), consistent with a larger induced gap, as expected for more transparent contacts. Also the zero-bias value of $dV/dI$ (with respect to $R_n$) is much smaller than in the previous sample.

That we are approaching the supercurrent regime is convincingly
shown by measurements of the differential resistance as function of
magnetic field (Figs.~5b,~c). At zero-bias, we observe an oscillating
behaviour of $dV/dI$ with maxima at $B = n \Delta B$ for $n = \pm 1,
\pm 2, \ldots$ (see dotted lines in Fig.~5b), and minima for $n = 0,
\pm 1.5, \pm 2.5, \ldots$ (with $\Delta B \approx 1.5$ mT). This
periodic behaviour is closely related to a Fraunhofer pattern that
is usually observed in the supercurrent regime of a Josephson
junction \cite{Tinkham}, and which shows the presence of a uniform current
through proximity-induced coherent states between the two Nb
contacts. The effective area ($A$) can be determined from the
periodicity of the $dV/dI$ oscillations, and is equal to $A =
\Phi_0/\Delta B \approx$ 1.3 $\mu$m$^2$ (where $\Phi_0=h/2e$ is the
flux quantum). The width of the HgTe stripe is 2 $\mu$m, giving an
effective length of 650 nm, which is approximately equal to the
distance between the centers of the Nb leads. This result clearly
shows that a precursor of the Josephson effect occurs in the
topological surface states of strained HgTe.

\begin{acknowledgments}


We thank C. Ames, P. Leubner, M. M\"{u}hlbauer, and C. Thienel
for fabrication and characterization of the HgTe layers, and
B. Trauzettel and P. Recher for useful discussions. This work was
financially supported by the German research foundation DFG
[SPP 1285 Halbleiter Spintronik, DFG-JST joint research program
'Topological Electronics', Emmy-Noether Grant, No. HA5893/3-1],
and the EU ERC-AG program [project 3-TOP].

\end{acknowledgments}

\end{document}